\documentstyle[12pt,aps,psfig]{revtex}

\tightenlines
\begin{document}
 
\draft

\title{Economics Mapping to the Renormalization Group \\
Scaling of Stock Markets}
\author{Enrique Canessa\thanks{E-mail: canessae@ictp.trieste.it}}
\address{The Abdus Salam International Centre for Theoretical
Physics, Trieste, Italy}
%\date{\today}
 
\maketitle
\begin{abstract}
\baselineskip=12pt

We make an attempt to map a simple economically motivated model for the
price evolution [J. Phys. A: Gen. Math {\bf 33}, 3637 (2000)] to the
phenomenological renormalization group scaling of stock markets.  This 
mapping gives insight into the critical exponents and the 
renormalization group predictions for the log-periodic oscillations
preceding some stock market crashes from the perspective of non-linear
changes in {\it `the level of stocks'}. \\

%\pacs{89.90.+n, 01.75+m, 02.50.+s}

\end{abstract}

\baselineskip=20pt

\section{introduction}

Several papers have appeared in recent years showing increasing
evidence that at least some market crashes are often anticipated by a
power law behaviour of the stock market index which fluctuates with
oscillations that are periodic in the logarithm of the time to crash
(see, for example, \cite{Sor96,Sor97,Fei96,Glu98,Van98,Joh00}).  From
these observations, it has been argued that there is a close relation
between the stock market crashes and the renormalization group (RG)
theory \cite{Sor96}.  Precursory  logarithm- (log-)periodic patterns 
can also emerge from percolation models by applying the cluster concept
to groups of investors acting collectively \cite{Sta98,Sta00}. 

Though the RG approach has shown to model remarkably well the stock
market time evolution and to predict the existence of a large price
crash, a possible universality for the real exponents quantifying the
observed behaviour in the market prices -which would allow to define a
crash- has not been yet established \cite{Sor97}.  Unlike in the case
of systems in thermodynamic equilibrium, there is no known underlying
Hamiltonian from which RG critical exponents could be deduced. In this
paper we propose a simplified dynamics for the price evolution and
made an attempt to map this dynamics to the RG predictions.  We show 
how the simplest, non-linear economics model proposed by the Author
\cite{Can00} may be mapped into the non-linear RG scaling of stock
markets in order to understand the critical exponents in terms of 
relevant economics variables such as the demand and supply of a 
commodity (or product).  As an illustrative example, we apply the
mapping to the NY Standard \& Poor (S\&P)500 index crash of Oct. 1987.

\section{Non-Linear Renormalization Group Generalization}

In analogy with RG theory, it is assumed that
the temporal variation of the stock market index $I(t)$ is related to future
 events at $t'$ by the transformations \cite{Sor96,Sor97}
\begin{equation} \label{eq:firstt}
x' = \phi(x)   \;\;\; ,
\end{equation}
\begin{equation}\label{eq:secondd}
F(x) \equiv I(t_c)-I(t)  = g(x) + \frac{1}{\mu} F\biggl(\phi(x)\biggl) \;\;\; ,
\end{equation}
where $x = t_{c}-t$.  $\phi$ is called the RG 
flow map and $\mu$ is a constant describing the scaling of $I$
upon the rescaling of $t$ in Eq.(\ref{eq:firstt}).
$F=0$ at the critical point $t_{c}$ (the time of a large crash),
$g(x)$ represents the non-singular part of the function
$F(x)$, which is assumed to be continuous, and $\phi(x)$ is assumed to
be differentiable.

An extension of these results to a more general RG approach begins by
considering that the solution of the RG Eq.(\ref{eq:secondd}), in
conjunction with Eq.(\ref{eq:firstt}) and the linear approximation
$\phi(x) = \hat{\lambda} x$, can be rewritten as
\begin{equation}\label{eq:ertf}
{dF(x) \over d\log x} =  \alpha F(x) \;\;\; ,
\end{equation}
This defines limiting power laws as $t \rightarrow t_{c}$.
Eq.(\ref{eq:ertf}) is then extended to include
corrections to the power law with log-periodicity by 
introducing the amplitude $B$ and phase $\psi$
of $F(x)~=~Be^{i\psi(x)}$.   The symmetry law used
is the phase shift that should keep the observable constant under a 
change of units \cite{Sor97}. This leads to postulate the following 
Landau expansion \cite{Gol92}:
\begin{equation}\label{eq:azepo}
{dF(x) \over d\log x} =  (\alpha + i \omega) F(x)
+ (\eta + i \kappa) |F(x)|^2 F(x) + {\cal O}(F^5)   \;\;\; .
\end{equation}
where $\alpha>0$, $\omega$, $\eta$ and $\kappa$ are real coefficients and
${\cal O}(F^5)$ represents higher order terms to be neglected. 

In terms of the amplitude and phase of $F$, Eq.(\ref{eq:azepo}) yields
\begin{equation}\label{eq:bpsi}
{\partial B \over \partial \log x} = \alpha B + \eta B^{3} + \cdots  \;\;\; ; \;\;\;
{\partial \psi \over \partial \log x} = \omega + \kappa B^{2} + \cdots 
\end{equation}
whose solutions are found to be
\begin{equation}\label{eq:soklj}
B^{2}  =  {({x \over x_0})^{2\alpha}
\over { 1 + ({x \over x_0})^{2\alpha}} } \; B_{\infty}^{2} \;\;\; ,
\end{equation}
\begin{equation}\label{eq:poiin}
\psi  = \omega \log ({x \over x_0}) +
{\Delta \omega \over 2 \alpha} \log \biggl( 1
+ ({x \over x_0})^{2\alpha} \biggl)  \;\;\; ,
\end{equation}
where $B_{\infty}^2 = \alpha / |\eta|$, 
$\Delta \omega = B_{\infty}^2 \kappa$ and $x_0$ is an arbitrary
coefficient characterizing the time scale.  These general forms lead to the
following solutions of the non-linear RG Eq.(\ref{eq:azepo}):
\begin{eqnarray}\label{eq:sixthhhh}
I\left( \tau \right) & =  & A_{1} + A_{2} {\left( \tau_c-\tau \right) ^{\alpha} 
  \over \sqrt{ 1 + \left( {\tau_c-\tau \over \Delta t} \right)^{2 \alpha}}} 
     \; \times \nonumber \\
  &  & \hspace{2cm} 
\left[ 1+ A_{3} \cos \biggl( \omega \log \left( \tau_c-\tau \right) 
+ {\Delta \omega \over 2 \alpha} \log \left( 1 + \left( {\tau_c-\tau \over \Delta
t} \right)^{2\alpha} \right) \biggl) \right] \;\;\; ,
\end{eqnarray}
where $\tau = t/\phi$, $\Delta t = x_{0}$ and $A_{i=1,2,3}$ are linear
variables.

\section{Non-linear Economics Model}

Within our economics model \cite{Can00}, only one stock of the
commodity is assumed and the market is considered competitive so it
self-organizes to determine the behaviour of the asset price $p$.  We
derive a dynamical price equation which results from the prevailing
market conditions in terms of the excess demand function
$E(p)=D(p)-Q(p)$, where $D$ and $Q$ are the demand and supply
functions, respectively.  In our description, asterisks ($^{*}$) denote
quantities in equilibrium and all variables are dimensionless. 

In a competitive market the rate of price increase usually is a functional 
of $E(p)$ such that $dp/dt \equiv f[ E(p) ]$ \cite{Cle84}.  Considering
that in general a commodity can be stored, then stocks of the commodity 
build up when the flow of output exceeds the flow of demand and vice-versa.
The rate at which {\it `the level of stocks'} $S$ changes 
can be approximated as $dS/dt = Q(p) - D(p)$.  Thus a price
adjustment relation that takes into account deviations of the stock
level $S$ above certain optimal level $S_{o}$ (to meet any demand reasonably
quickly) is given by
\begin{equation}\label{eq:stock}
\frac{dp}{dt} = -\gamma \frac{dS}{dt} + \lambda (S_{o}-S)  \;\;\; ,
\end{equation}
where $\gamma$ ({\it i.e.}, the inverse of excess demand required
to move prices by one unity \cite{Bou98}) and $\lambda$ are positive
factors.  For $\lambda > 0$, prices increase when stock levels are 
low and raise when they are high (with respect to $S_{o}$). 
When $\lambda = 0$, the price adjusts at a rate proportional
to the rate at which stocks are either raising or running down. 

For all asset prices $p(t)$, non-linear forms for the quantities $D$ 
demanded and $Q$ supplied are postulated such that
\begin{eqnarray}\label{eq:dq}
D(p) & = &  d^{*} + d_{o} [ \; 1 - \frac{\delta^{2}}{2!}(p-p^{*})^{2} + 
       \dots \;] (p-p^{*}) \;\;\; ,
\nonumber \\
Q(p) & = &  q^{*} + q_{o} [ \; 1 - \frac{\delta^{2}}{2!}(p-p^{*})^{2} + 
       \dots \;] (p-p^{*}) \;\;\; , 
\end{eqnarray}
where $d_{o}$, $q_{o}$ and $d^{*}=D(p^{*})$, $q^{*}=Q(p^{*})$ are arbitrary
coefficients (related to material costs, wage rate, {\it etc}), 
$p^{*}=p(t^{*})$ is an equilibrium price and $\delta < 0$ is our order
parameter as discussed in \cite{Can00}.  Expansion terms ${\cal O}(5)$ are 
here neglected. 

Considering the simplest, complete economics model as in \cite{Cle84}, 
we assume that $S_{o}$ depends linearly on the demand; {\em e.g.},
$S_{o} = \ell_{o} + \ell D$, with $\ell_{o}$ a constant and $\ell$
satisfying 
\begin{equation}\label{eq:ell}
\ell \equiv \frac{\gamma \beta_{o}}{\lambda d_{o}} \;\;\; ,
\end{equation}
where $\beta_{o} \equiv q_{o}-d_{o}$.
We have shown that only this condition can lead to solutions 
of the dynamical price equation in real space \cite{Can00}.
Therefore, in equilibrium (where $\frac{dp}{dt}|_{p^{*}}=0$ and
$\frac{dS}{dt}|_{S^{*}}=0$, so that demand equals supply and $S=S_{o}$),
we obtain $d^{*}-q^{*} = 0$ and 
$S^{*} = \ell_{o} + \ell ( d^{*} + d_{o} p^{*} )$.

After some algebra, the second derivative of the price adjustment 
Eq.(\ref{eq:stock}) of one commodity can be approximated as 
\begin{equation}\label{eq:dyn}
\frac{d^{2}p}{dt^{2}} \; \approx \;  - \lambda \beta_{o} (p-p^{*}) +
   \frac{\delta^{2} \lambda \beta_{o}}{2}(p-p^{*})^{3}  \;\;\; .
\end{equation}
For $\delta \ne 0$ and
$[\lambda \beta_{o},\delta^{2} \lambda \beta_{o}/2] > 0$,
it has the well-known kink solutions
\begin{equation}\label{eq:kink}
p(t) =   p^{*} + \frac{\sqrt{2}}{\delta} \; \tanh
  \biggl( \sqrt{\frac{\lambda \beta_{o}}{2}} \; (t-t^{*}) \biggl) \;\;\; ,
\end{equation}
such that $\beta_{o}$ is positive.
As in a competitive market economy the demand for a 
commodity fall when its price increases, then it is reasonable to
assume $d_{o} < 0$ in Eq.(\ref{eq:dq}).  And as the price raises, the supply
usually also increases; hence in general one also assumes $q_{o} > 0$.  
These conditions yield $\beta_{o} > 0$ as required and also $d_{o}\ell >0$.

\section{The Mapping}

We show next how a mapping can be established in order to identify the
real, phenomenological RG critical exponents $\alpha$ and $\eta$ in
terms of our non-linear economics model variables.
From a comparison between Eqs.(\ref{eq:bpsi}) and (\ref{eq:dyn}) we 
identify the following relation between the expansion terms 
\begin{eqnarray}\label{eq:map1}
\alpha B & \rightleftharpoons & - \; \delta \beta_{o} (p -p^{*}) \;\;\; . 
\end{eqnarray}
It is from this simple mapping that we can made an attempt to understand 
RG modelling of stock markets and from it to analyse and predict financial 
crashes in analogy to critical points studied in physics with log-periodic
correction to scaling \cite{Sor97}.

If for $t \rightarrow t^{*}$ we approximate 
$\sinh \biggl( \sqrt{\frac{\lambda \beta_{o}}{2}} \; (t-t^{*}) \biggl) 
\approx \sqrt{\frac{\lambda \beta_{o}}{2}} \; (t-t^{*})$ such that
\begin{equation}\label{eq:sinh}
  \tanh \biggl( \sqrt{\frac{\lambda \beta_{o}}{2}} \; (t-t^{*}) \biggl) \approx 
  \frac{\sqrt{\frac{\lambda \beta_{o}}{2}} \; (t-t^{*})}
  {\sqrt{1+\frac{\lambda \beta_{o}}{2} \; (t-t^{*})^{2}}} \;\;\; ,
\end{equation}
then using the mapping in Eq.(\ref{eq:map1}) and the solutions
for $B$ and $p$ given by Eqs.(\ref{eq:soklj}) and (\ref{eq:kink}), respectively, we obtain
\begin{eqnarray}\label{eq:alphaeta}
\alpha & \rightleftharpoons & \lim_{t \rightarrow t_{c} \rightleftharpoons t^{*}} \;
          \frac{ \log \biggl( (t^{*}-t)/\sqrt{2/\lambda \beta_{o}} \biggl) }
               { \log (x/x_{o}) } \;\; \rightarrow \; 1   \;\;\; , \nonumber \\
|\eta| & \rightleftharpoons & \frac{\delta^{2}}{ 2 \lambda^{2} \beta_{o}^{2}} \;\;\; , 
\end{eqnarray}
such that, as before $x=t_{c}-t$ and $x_{o}=\Delta t$, and 
$\Delta t \rightarrow \sqrt{ 2/\lambda \beta_{o}}$.  The above $\alpha$ is consistent
with the definition of critical exponents \cite{Gol92}.

Using these mappings for $\alpha \rightarrow 1$ and $\eta$, it is 
straightforward to show that they also relate the second series 
expansion terms between Eqs.(\ref{eq:bpsi}) and (\ref{eq:dyn}), namely:
$\eta B^{3}  \rightleftharpoons  \frac{\delta^{2} \lambda \beta_{o}}{2}(p-p^{*})^{3}$
provided that $\delta < 0$.  Hence, in terms of our non-linear economics
model variables, we find the following extended solutions in analogy with
the non-linear RG framework: 
\begin{eqnarray}\label{eq:map2}
I\left( t \right) & =  & A_{1} + A_{2} {\left( t^{*}-t \right) 
  \over \sqrt{ 1 + \left( { t^{*}-t \over \Delta t} \right)^{2}}}
     \; \times \nonumber \\
  &  & \hspace{2cm}
\left[ 1+ A_{3} \cos \biggl( \omega \log \left( t^{*}-t \right)
+ {\Delta \omega \over 2} \log \left( 1 + \left( { t^{*}-t \over \Delta
t} \right)^{2} \right) \biggl) \right] \;\;\; ,
\end{eqnarray}
with $\tau \rightleftharpoons t$, $\tau_{c} \rightleftharpoons t^{*}$,
$\Delta \omega  \rightleftharpoons 2 (\lambda \beta_{o}/\delta )^{2} \kappa$, 
and $\Delta \tau \rightleftharpoons \sqrt{ 2/\lambda \beta_{o}}$.

\section{Discussion}

As an example, we apply next the present mapping to the S\&P500 index. 
In Fig. 1 we show the fit of Eq.(\ref{eq:map2})  to the time dependence
of the logarithm of the NY S\&P 500 index from Jan 1985 to the October
1987 crash.  The parameters used in this illustrative curve (full line)
are: $A_{1}=5.79$, $A_{2}=-0.32$, $A_{3}=0.059$, $w=6.47$, $\Delta
t=2.29$, $\Delta \omega=15.42$ and $t^{*}=87.70$ decimal years (with
$rms=0.02$).  The parameter values used for the best fit of
Eq.(\ref{eq:sixthhhh}) (dotted lines in the figure) are those found in
\cite{Sor97}. 

Similarly to the non-linear RG scaling results, we see that the general
trend of the S\&P 500 data is also reproduced by the mapping of our
economics model in the limit $t \rightarrow t^{*}$ so that $\alpha
\rightarrow 1$ as deduced from Eq.(\ref{eq:alphaeta}).  This is to be
expected due to the oscillations that are periodic in the logarithm of
the time to crash appearing in both Eqs.(\ref{eq:sixthhhh}) and
(\ref{eq:map2}).

%%%%%%%%%%%%%%%%%%%%%%%%%%%%% Figure 1 and Caption
\begin{figure}
\centerline{\hbox{ \psfig{figure=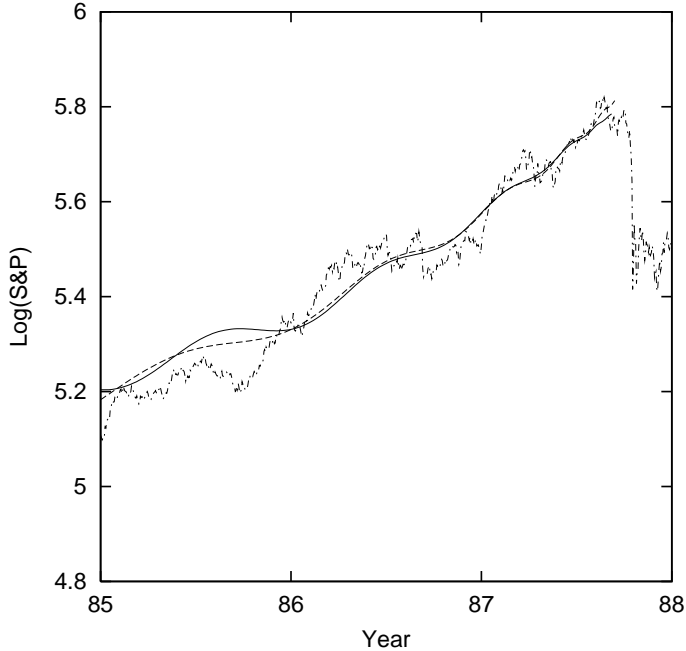,width=5in,angle=270}} }
\vspace{0.6cm}
\caption{\narrowtext Time dependence of the logarithm of the NY S\&P 500
index from Jan 1985 to the October 1987 crash.  The full line
curve is the fit of Eq.(\ref{eq:map2}) for a year scale of
$2.29$ years and the dotted lines curves the fit of Eq.(\ref{eq:sixthhhh})
with $\Delta t=11$ years. }
\label{fig1}
\end{figure}
%%%%%%%%%%%%%%%%%%%%%%%%%%%%%

The modulation of the logarithm frequency does not changes
significatively when one approaches the critical point $t_{c}
\rightleftharpoons t^{*}$.  On relatively short time scales spikes are
not accounted by both equations due to the complex self-organizing
phenomena in stock markets other than the one analysed here.  The
log-periodic structure found prior to crashes implies the existence of
a hierarchy of time scales \cite{Sor97}.  The choice of the parameters
$A_{i=1,2,3}$ is still empirical in both cases. 

Our mapping may give insight into the nature of market crashes from the
new perspective of the demand $D$ and supply $Q$ of a commodity.  Our
non-linear expressions for $D$ and $Q$ of Eq.(\ref{eq:dq})  are plotted
in Fig. 2 and are justified as follows: when $|\delta p| << 1$, these
functions display similar behaviour to the (commonly used) linear
$p$-dependence for $D$ and $Q$.  Even more important, they depict the
fact that as price falls, the quantity demanded for a commodity can
increase in agreement with one of the basic principles of economy.  On
the other hand, our choice for $Q$ (with $q_{o}>0$) also follows the
typical behaviour observed in a competitive market (where no individual
producer can set his own desired price).  That is, the higher the
price, the higher the profit, then the higher the supply (see
\cite{Can00} for a more extensive discussion).

%%%%%%%%%%%%%%%%%%%%%%%%%%%%% Figure 2 and Caption
\begin{figure}
\centerline{\hbox{ \psfig{figure=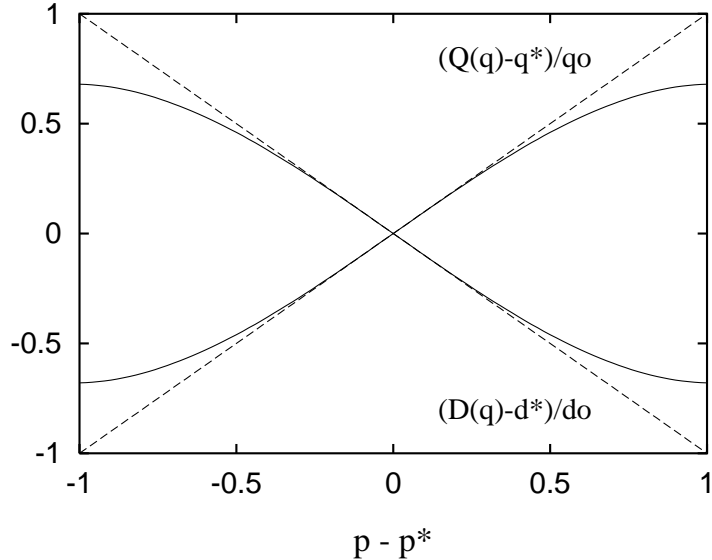,width=4in,angle=270}} }
\vspace{1cm}
\caption{\narrowtext Non-linear forms for the demand and supply functions
of Eq.(\ref{eq:dq}) with $q_{o}>0$, $d_{o}<0$, $|\delta|=0.8$ (full lines) 
and $\delta=0$ (dotted lines).}
\label{fig2}
\end{figure}
%%%%%%%%%%%%%%%%%%%%%%%%%%%%%

We have related the critical exponents $\alpha$ and $\eta$ to the
relevant variables of our non-linear economics model based on observed
laws for $D$ and $Q$ \cite{Can00}.  As $t \rightarrow t^{*}$ we have
identified $\alpha \rightarrow 1$ whereas the $\eta$ exponent given in
Eq.(\ref{eq:alphaeta}) is found to depend on the order parameter
$\delta < 0$, relating $D$ and $Q$ of a commodity as in
Eq.(\ref{eq:dq}) in conjunction with the economics model factors
$\lambda$ and $\beta_{o}$ under the limiting constraints of
Eq.(\ref{eq:sinh}). 

Since our $\Delta t$ coefficient also depends on $\lambda$ and
$\beta_{o}$, these factors drive the observed effects for $t/t_{c} >>
1$, where there is a saturation of the function $I(t)$, and for $t
\rightarrow t_{c}$, where the log-frequency shifts from ${\omega +
\Delta \omega \over 2\pi}$ to ${\omega \over 2\pi}$.  In economics
terms, these features are directly related to the temporal adjustments
of {\it `the level of stocks'} $S$ as given in Eq.(\ref{eq:stock}). 

The concept of a certain optimal level of stock is well-known
in economics theory about stocks \cite{Cle84}.  
Planning ahead to have suitable {\it `level of stocks'} is essential. 
If production had to be stopped every time a company ran out of raw 
materials, the time wasted would cost a fortune.

Indeed stocks are held for a variety of reasons.  There may be stocks of raw
materials ready for production, stocks of work-in-progress ({\it e.g.},
production parts)  or stocks of finished goods.  Whichever they are it
is vital for a company to control {\it `the level of stocks'} very
carefully.  Too little and they may run into production problems, but
too much and they have tied up money unnecessarily.  Low {\it `level of
stocks'} -say 10\%- would certainly be adequate if production levels
could be maintained during the years.  Usually {\it `the level of
stocks'} needs to be adjusted as the marketing year progresses. 
Stocks are considered to be current assets because they can be
converted into cash reasonably quickly.  On the other hand, producers 
can also carry some stocks surplus as a way to speculate on prices. 

\section{Conclusions}

We have shown that the present economics mapping to the RG scaling of
stock markets reasonably allows to reproduce the S\&P 500 index trends
in the vicinity of the time of crash and also to predict the existence
of a crash as in the RG model due to corrections to the power law with
log-periodicity but such that $\alpha \rightarrow 1$. The main point of
this work follows that of RG model.  That is, the underlying cause of
the crash must be searched years before by looking at the progressive
accelerating ascent of the market price.  However, our formalism
differs in that we have a shorter year scale of about $\Delta t=2.29$
years compared to the fitting of $\Delta t=11$ years as reported in
\cite{Sor97} in which these cooperative phenomena are progressively
being constructed.  In this period of time, one should also look for
the appearence of non-linearities in the behaviour of the demand and
supply functions (or, alternatively, in {\it `the level of stocks'}) of 
the commodities prior to the crash as recognized here. 

\vspace{1cm}
{\it Acknowledgements:} The author thanks the Organizers of the 2nd European
Conference on {\it `Applications of Physics in Financial Analysis'} held at
Li{\`e}ge, Belgium (July 2000), and the European Physical Society for financial support.

\end{document}